\documentclass[a4paper,10pt]{article}
\usepackage[utf8]{inputenc}
\usepackage{amssymb}

\title{ Generalized Uncertainty Principle as a Consequence  of the Effective Field Theory}
\author{Mir Faizal$^{1,2}$,  Ahmed Farag Ali$^{3,4}$, Ali Nassar$^5$ \\\\
\\
$^{1}$Irving K. Barber School of Arts and Sciences,
\\ University of British
Columbia - Okanagan \\
  Kelowna,  British Columbia V1V 1V7, Canada.\\
$^2$Department of Physics and Astronomy, \\  University of Lethbridge, Lethbridge \\  Alberta T1K 3M4, Canada. \\
 $^3$Department  of Physics, Faculty of Science, \\
 Benha University, Benha, 13518, Egypt.\\
 $^4$Netherlands Institute for Advanced Study, \\ Korte Spinhuissteeg 3, 1012 CG Amsterdam. Netherlands\\
$^5$Department of Physics, University of Science and Technology, \\Zewail City of Science and Technology,
 12588 Giza, Egypt}
\date{}

\begin{document}

\maketitle
\begin{abstract}
 We will demonstrate that the  generalized uncertainty principle exists because of the derivative expansion in the effective field theories. This is because  in the framework of the effective field theories, the minimum measurable length scale  has to be integrated away to obtain the low energy effective action. We will analyze the deformation of a massive free scalar field theory by the generalized uncertainty principle, and demonstrate that the minimum measurable length scale corresponds to a second  more massive scale in the theory, which has been integrated away. We will also analyze CFT operators dual to this deformed scalar field theory, and observe that scaling of the new CFT operators indicates that they are dual  to this more massive scale in the theory. We will use holographic renormalization to explicitly calculate the renormalized boundary action with counter terms for this scalar field theory deformed by generalized uncertainty principle, and show that the generalized uncertainty principle  contributes to the matter conformal anomaly.
\end{abstract}

\section{Introduction}

It is a universal prediction of almost all approaches to quantum gravity, that there is a minimum measurable length scale  and it is not possible to make measurements below that scale. In perturbative string theory, the string length scale acts as the minimum measurable length scale. This is because in perturbative string theory, the smallest probe that can be used for analyzing any region of spacetime is the string, and so, spacetime not be probed at length scales below string length scale \cite{a}. The existence of a minimum length scale  in loop quantum gravity  turns the big bang into a big bounce \cite{z1}.  The generalized uncertainty principle has also been obtained from quantum geometry \cite{quan}. This has been done by taking into account the existence of an upper bound on the acceleration of massive particles \cite{acce}-\cite{acce1}. So, the generalized uncertainty principle can also be motivated from a deformation of the geometry of spacetime by a constraint on the
maximal acceleration of massive particles. It may be noted that the deformation of spacetime has also been  analyzed using conformal transformations \cite{conf}. The energy needed to probe spacetime at length scales smaller than Planck length is more than the energy required to form a black hole in that region of spacetime. So, spacetime  cannot be probed below the Planck scale, as this will lead to the formation of a mini black holes, which will in turn restrict the measurement of any phenomena below the Planck scale. Thus, the existence of a minimum measurable length scale can also be inferred from black hole  physics \cite{z4}-\cite{z5}. On the other hand, the existence of a minimum measurable length scale is not consistent with the usual Heisenberg uncertainty principle. This is because according to the usual Heisenberg uncertainty principle, the length can be measured to arbitrary accuracy if the momentum is not measured. To incorporate the existence of a minimum measurable length scale in the Heisenberg uncertainty principle, one needs to modify it to a generalized uncertainty principle. However, as the Heisenberg uncertainty principle is related to the Heisenberg algebra, the deformation of the Heisenberg uncertainty principle also deforms the Heisenberg algebra \cite{7q}-\cite{c}.

The deformed  Heisenberg algebra in turn deforms the coordinate representation of the momentum operator \cite{7q}-\cite{c}. This corrects all quantum mechanical systems, including the first quantized equations of  a field theory \cite{skdj}. In fact, a covariant version of this deformed algebra is used to deform the field theories  \cite{di}, and this covariant deformation is consistent with the existence of a minimum measurable time \cite{ti}, apart from being consistent with the existence of a minimum measurable length. The gauge theories corresponding to such a deformed field theory has also been studied \cite{d}-\cite{h}. In this paper, we will analyze some  theoretical aspects of such a deformed field theory. We will also use  the holographic principle to understand the boundary dual of such a deformed field theory. The holographic principle states that the gravitational degrees of freedom  in a region are encoded in the boundary degrees of freedom of that region. One of the most successful realization of the holographic principle is the gauge/gravity duality also known as the $AdS/CFT$ correspondence \cite{1}-\cite{1c}. This duality relates  type IIB string theory on $AdS_5 \times S^5$ to
$\mathcal{N}=4$ super-Yang-Mills theory on  its conformal  boundary. It may be noted that even though the full string theory on  $AdS_5 \times S^5$ is not understood, this duality can be used to map the weakly  coupled limit of  string theory to the strongly coupled
gauge theory \cite{5}. In fact, it can also be used to map a strongly coupled limit of the string theory to the weakly coupled limit of the gauge theory \cite{5a}. Thus, this duality can be used for analyzing  the strongly coupled limit of the gauge theory by analyzing weakly coupled limit of the string theory. Since the weak coupling limit of the string theory can be approximated by  ten dimensional supergravity, this duality is usually used to map the ten dimensional supergravity on $AdS_5 \times S^5$ to $\mathcal{N} = 4$ super-Yang-Mills theory on its conformal boundary. Furthermore, to suppress the loop contributions of the ten dimensional supergravity one takes the large $N$ limit of the gauge  theory. It may be noted that the UV divergences of the correlation functions on the gauge theory side need to be renormalized. However, these UV divergences are related to the IR divergences on the gravitational side of the duality. The IR divergences on the gravitational side are the same as near-boundary effects, and so, they can be  dealt with by using holographic  renormalization \cite{4}-\cite{4a}. This is because  the cancellation of the UV divergences does not depend on the IR physics, and this in turn  implies that the holographic renormalization should only depend on the near-boundary analysis.

It may also be noted that even though the $AdS/CFT$ conjecture has been mostly used in the context  of string theory, this conjecture is actually a more general conjecture. In fact, the $AdS/CFT$ conjecture has also been used for analyzing Rehren duality also known as algebraic holography \cite{a1}-\cite{b1}. The Rehren duality establishes the correspondence between an ordinary scalar
field theory on $AdS$ and a suitable conformal field theory on its boundary. In  Rehren duality a   space like wedge  in $AdS$ is mapped to its intersection with the boundary \cite{c1}. This sets up a bijection between the set of all wedges in the bulk and the set
of all double-cones on the boundary. In fact, this bijection maps spacelike related bulk wedges to spacelike related boundary
double-cones. Now for a net of local algebras on the bulk the Rehren duality defines  a net of local algebras on the boundary. This is done  by identifying the algebra for a given boundary double-cone with the bulk wedge algebra which restricts to it. In fact, another approach that  relates a ordinary scalar field theory in the bulk to the conformal field theory on its boundary is the boundary-limit holography \cite{d1}. Thus, the main idea behind $AdS/CFT$ conjecture has wider applications than  relating type IIB string theory on  $ AdS_5 \times S^5$ to the $\mathcal{N} =4$ super-Yang-Mills theory on its boundary.

Using this as a motivation, we will analyze the boundary dual of a scalar field theory with higher derivative corrections in the bulk. Higher derivative corrections to the scalar field theory have been predicted from discrete spacetime \cite{1q}, spontaneous symmetry breaking of Lorentz invariance in string field theory \cite{2q}, spacetime foam models \cite{3q}, spin-network in loop quantum gravity \cite{4q}, non-commutative geometry \cite{5q}, Horava-Lifshitz gravity \cite{6q}, and the existence of minimum length \cite{7q}. In fact, the existence of the string length scale  also produces   higher derivative corrections to the  low energy phenomena \cite{0000a1}-\cite{0000a2}. Aspects of higher derivative terms have been investigated in cosmological inflation in \cite{greene}-\cite{greene1}. In fact, motivated by the existence of a minimum length in string theory,  higher derivative corrections to the scalar field theory in $AdS/CFT$ has been recently analyzed \cite{st}. As we are analyzing low energy effective phenomena, these higher derivative corrections are also expected to occur due to the derivative expansion in the   effective field theory \cite{effe}-\cite{effe1}. In this paper, we will analyze a scalar field theory deformed by generalized uncertainty principle, and observe that it contains higher derivative corrections.  We will also analyze the physical meaning of these higher derivative terms. It has been suggested that the  high energy excitations in the bulk will correspond to the CFT operators scaling as  $\Delta_1\sim N^{2/3}$ in five dimensions, or $\Delta\sim N^{1/4}$ in ten dimensions \cite{1c}. We observe that the deformation of the scalar field theory by the generalized uncertainty principle in the bulk  produces CFT operators with this scaling property on the boundary. This implies that the deformation produced by the generalized uncertainty principle actually correspond to high energy excitation in the bulk, as was expected from effective field theory.

\section{Deformed Field Theory}

In this section, we will analyze the deformation of a massive scalar field theory on $AdS$ by the generalized uncertain principle
\cite{d}-\cite{h}. Furthermore,  it will be demonstrated that the
higher derivative corrections obtained from the generalized uncertainty principle will be exactly the same as the correction generated from a derivative expansion in the light of effective field theories \cite{effe}-\cite{effe1}. The existence of minimum measurable length causes the following deformation of the uncertainty principle, and for a simple one dimensional system it can be written as $\Delta x \Delta p =   [1 + \beta (\Delta p)^2]/2 $ \cite{7q}-\cite{c}. Here   
$\beta = \beta_0 \ell_{Pl}^2 $ and  $\beta_0$ is a constant normally assumed to be of order one, and this corresponds to taking the
Planck length $\ell_{Pl} \approx 10^{-35}~m$ as the minimum length scale. However, it is possible to take the minimum measurable length scale as an  intermediate length scale $\ell_{Inter}$, which is between the Planck length scale  and electroweak length scale. In this case, the constant $\beta_0$ will be given by $\beta_0\approx \ell_{Inter}^2/ \ell_{Pl}^2$ \cite{7q}.  It may be noted that this will change the value of $\beta$, and as we will demonstrate that $\beta$ acts as another mass scale in the theory, this will change the value of that mass scale. However, in this paper,  we will fix the value of $\beta_0 \approx 1$ by taking the Planck scale as the minimum measurable length scale. This deforms the Heisenberg algebra, as the Heisenberg algebra is closely related to the Heisenberg uncertainty principle. The deformed  Heisenberg algebra in any dimension can be written as
\begin{equation}
[x^i, p_j ] =
i  [\delta_{j}^i + \beta p^2 \delta_{j}^i + 2 \beta p^i p_j].
\end{equation}
The coordinate representation of the deformed momentum, which is consistent with this algebra  is
\cite{d}
\begin{equation}
 p_\mu =- i  \partial_\mu  (1 -  \beta \partial^\nu \partial_\nu ).
\end{equation}

We will analyze such a deformation of a free scalar field theory  on $AdS$. The $AdS$ metric  can be written as
\begin{equation}
ds^2=G_{MN}dx^M dx^N=  L^2 z^{-2} [dz^2 + \delta_{\mu\nu}dx^\mu dx^\nu].
\end{equation}
The Laplacian on $AdS$  is given by
\begin{equation}
\Box = z^{d+1}\partial_z(z^{-d} \partial_z) + \Box_0,
\end{equation}
where $\Box_0=\delta^{ij}\partial_i \partial_j$.
Thus, a covariant version of the deformed momentum on $AdS$ can be written as \cite{st}
\begin{equation}
 p_M =- i  \nabla_M  (1 -  \beta \Box  ).
\end{equation}
It may be noted that the original momentum on $AdS$ was $\tilde p_{M } = -i \nabla_M $, so the effect of generalized uncertainty principle
is that it deforms $\tilde p_M \to   p_M$ and  this  deforms
$ \nabla_M  \to   \nabla_M  (1 -  \beta \Box  ),
$ which in turn deforms (to the leading order in $\beta$) $\Box \to \Box -  \beta \Box^2 $.
The action of the original free massive scalar field theory can be written as
\begin{equation}
S=  \frac{1}{2} \int d^{d+1}x \sqrt{g}\big[   G^{MN} \nabla_M \Phi \nabla_N \Phi +    m^2 \Phi^2\big].
\end{equation}
The equation of motion for this free massive scalar field theory on $AdS$,
can be written as,
\begin{equation}
\Big(\Box -m^2 \Big)\Phi=0.
\end{equation}
This equation will get deformed by the generalized uncertain principle as \cite{st}
\begin{equation}
\Big(\Box -m^2-   \beta \Box ^2\Big)\Phi=0.\label{1b}.
\end{equation}
Now we can write the action for the deformed   scalar field theory,  which can produce Eq. (\ref{1b})
as its equation of motion,
\begin{equation}
S=  \frac{1}{2} \int d^{d+1}x \sqrt{g}\big[  G^{MN} \nabla_M \Phi \nabla_N \Phi + 2 \beta
 \Box \Phi\Box \Phi + m^2 \Phi^2\big].
\end{equation}
It may be noted that such higher derivative terms  will modify the propagator from $1/ (p^2 + m^2)$ to $1/( p^2 + m^2 + \beta p^4)$, and thus there will be additional poles. This propagator can be written as a sum of two propagators, $A_1/ (p^2 + m_1^2) + A_2/(p^2 + m_2) $, and
as it is possible for one of these  propagator to have a negative sign, such a theory can contain Ostrogradsky ghost \cite{aq}-\cite{bq}. In general, there are several problems with such higher derivative theories, and several different solutions have been proposed to deal with them. In fact, such theories  can be non-unitary and contain  negative norm states, which would produce negative probabilities. However, it is possible to use the Euclidean formalism, and trace over a certain field configuration in the final state \cite{hw}-\cite{hwhw}. The theory thus obtained is still non-unitary, but it does not contain negative norm states, and hence it does not produce negative probabilities. There are several other ways to deal with such higher derivative terms. It may be noted that Lee-Wick theories are  higher derivative  field theory, which are unitary \cite{w1}-\cite{lw}. So, we could also use the Lee-Wick formalism to analyze this deformed field theory. Thus, we can introduce an Lee-Wick field, and this  Lee-Wick field will correspond to a more massive mode. Then this theory would be unitary, if the  Lee-Wick field decays. This can occur by imposing suitable boundary condition, such that there are no outgoing exponentially growing modes. Even though such a boundary condition violates causality, in a Lee-Wick theory  such a violation only occurs at microscopic scales.  It has been argued that the macroscopic violation of causality does not occur in a Lee-Wick theory \cite{w1}-\cite{lw}. It may be noted that the a Lee-Wick scalar field theory, which had the same form as a  scalar field theory deformed by the generalized uncertainty principle, has been already analyzed using this formalism \cite{wl}. In fact,  it has been argued that component field theories obtained from a nonanticommutative deformation of supersymmetric field theory are Lee-Wick field theories \cite{nona}. It has also been demonstrated that this nonanticommutative deformation of a supersymmetric field theory is similar to  deformation produced  by the generalized uncertainty principle \cite{nona1}. So, it can be argued that the deformation of a scalar field theory by the generalized uncertainty principle can be analyzed using the Lee-Wick formalism.

It may be noted that such negative norm states even occur in the usual gauge theories, due to the gauge symmetry. However, due to the BRST symmetry, the Kugo-Ojima criterion can be used to remove these negative norm states from a usual gauge theory \cite{brst}- \cite{brst1}. So, it is expected that even such negative norm states can be removed by using a subsidiary condition in higher derivative theories. It has been argued that it might be possible to develop  such formalism to remove negative states from higher derivative theories \cite{r1}-\cite{j5}. In fact,  such higher derivative terms can also be analyzed using various other approaches \cite{b1}-\cite{b2}. Most of these approaches are also based on some superselection rule or some subsidiary condition which are used to remove the undesirable ghost states. In this paper, we will not be analyzing the unitarity of the higher derivatives terms in this deformed  field theory. We would be analyzing the relation between  the generalized uncertainty principle and effective field theories.
However, it is important to note that such higher derivative terms can be consistently handled using different approaches.

In the framework of effective field theories  \cite{effe}-\cite{effe1},
 for a given mass dimension, we have to include all terms, when performing the derivative expansion of the effective action. However, as
 we started from a free scalar field theory, the theory has to remain free at all scales. So, we cannot include higher   powers of fields
 in the derivative expansion of the effective field theory. Thus, we can only include higher order derivative terms in the effective action of this theory.
 The requirement of the theory to be Lorentz invariant further restricts the form of these higher order terms that can be added to such an effective action.
 In fact, to the first order,  the effective field theory action, satisfying these constraints,  is given by
\begin{equation}
S=  \frac{1}{2} \int d^{d+1}x \sqrt{g}\big[ G^{MN} \nabla_M \Phi \nabla_N \Phi + \frac{2}{M ^2}
 \Box \Phi\Box \Phi + m^2 \Phi^2\big].\label{1a}
\end{equation}
where $M^2$ is the scale which has been integrated out, and for theories based on generalized uncertainty principle, this has to be equal to
the minimum measurable length scale. In fact, if we identify $M^2 = \beta^{-1}$, we can observe that the action obtained 
from the derivative expansion in effective field theory is identical to the deformed action obtained from the  generalized uncertainty principle.
This is physically expected as generalized uncertainty principle is obtained by incorporating a minimum measurable length scale in the theory. In other words,
the theory is not defined below a certain length scale. However, if we look at this situation using effective field theory, we will have to integrate the
modes below that minimum length scale. Hence, both the effective field theories and generalized uncertainty principle seem to be based on the same physical principle,
i.e.,   not to make  measurements below a   certain length scale. This also suggests how we should handle  the deformation of a field theory by generalized uncertainty
principle. So, the parameter $\beta$ should be viewed as a perturbation parameter, and we should make a derivative expansion of the field theory.  This will act as a
new mass scale in the theory. However, as long as we are dealing with the infrared limit
of the theory, and the energy used to probe the theory will be small compared to this
mass of this mode, the theory will be well defined. This is the standard way to deal with such higher derivative terms in the framework of effective field theories
\cite{effe}-\cite{effe1},
and this is the framework we will use in this paper. The main interesting result  here is that the generalized uncertainty principle can be obtained as a consequence
of the derivative expansion of a free scalar field theory. This is because if we started from
Eq. (\ref{1a}), and made the identification, $M^2 = \beta^{-1}$,  the equation of motion  we would obtain would be the   equation of motion  deformed
  by the generalized uncertainty principle. In this paper, we will analyze the consequence of this correspondence further. In fact, we will use the $AdS/CFT$ correspondence
  to analyze the boundary dual to such a deformed field theory in the bulk.

In order to analyze this theory using the $AdS/CFT$ correspondence, we need to find an explicit expression for the boundary action.
The boundary action corresponding to this deformation can also be written as
\begin{equation}\label{boundaryaction2}
S_b=\frac{1}{2} \int d^{d+1}x \sqrt{\gamma}\,  n^\mu \Big[
\Phi\partial_\mu\Phi+\beta  \Big(\partial_\mu\Phi \Box\Phi-\Phi\partial_\mu\Box\Phi \Big)
 \Big],
\end{equation}
where $\gamma$ is the boundary metric.
Now using the Fourier transform of the scalar field  $\phi (z, x)$ in $x^\mu$ coordinates,
and  the ansatz $f_k(z)\approx z^\Delta$, for solutions close to the boundary, the solutions to this
equation can be written as,
\cite{h},
\begin{equation}\label{massVscd}
\beta^{-1} L^2\Delta (\Delta-d)-m^2\beta^{-1} L^4-\Delta^4+2d\Delta^3-d^2\Delta^2=0.
 \end{equation}
It may be noted that for the massless case, $m =0$, the roots have a simple structure,
\begin{equation}
\Delta_1=0, \hspace{0.4 in} \Delta_2=d, \hspace{0.4 in} \Delta_{3,4}={1 \over 2}[d\pm\sqrt{d^2+ 4\beta^{-1} L^2}].
\end{equation}
 These roots are very similar to that of mass deformations (scalar field mass terms), if we replace $m\rightarrow \beta^{-1}$.
 In fact, we can write the solution for the general massive case as follows,
\begin{eqnarray} \Delta_{1,2}&=&{1 \over 2}\left[d\pm\sqrt{d^2+ 2\beta^{-1} L^2+2\beta^{-1/2} L^2\sqrt{\beta^{-1}-4m^2}}\right], \nonumber\\
\Delta_{3,4}&=&{1 \over 2}\left[d\pm\sqrt{d^2+ 2\beta^{-1} L^2-2\beta^{-1/2} L^2\sqrt{\beta^{-1}-4m^2}}\right], \end{eqnarray}
For reasons that will be clear soon, let us call;
\begin{eqnarray} M_{+}&=&{ 1 \over 2}[\beta^{-1}+\beta^{-1/2}\sqrt{\beta^{-1}-4m^2}], \nonumber\\
M_{-}&=&{ 1 \over 2}[\beta^{-1}-\beta^{-1/2}\sqrt{\beta^{-1}-4m^2}], \end{eqnarray}
which will leave the four roots in even simpler form
\begin{eqnarray} \Delta_{1,2}&=&{1 \over 2}\left[d\pm\sqrt{d^2+ 4 M_{+}^2L^2}\right], \nonumber\\
\Delta_{3,4}&=&{1 \over 2}\left[d\pm\sqrt{d^2+ 4M_{-}^2L^2}\right]. \end{eqnarray}

Now one has to notice that $M_{+}= \beta^{-1/2} + O(\beta)$ and  $M_{-}= m +O(\beta)$,
which say that effectively we have two usual mass scaling dimensions.
Our solution in the bulk reads
\begin{equation} \Phi(x,z)\sim z^{\Delta_2}\, A(x)+ ....\end{equation}
where $\Delta_2={1 \over 2}\left[d-\sqrt{d^2+ 4 M_{+}^2L^2}\right]$ is the smallest root,
which leads the behavior of the solution near the boundary (close to $z=0$).
The connection between $\Phi$ and the CFT operator
is through the source term in the boundary action,
which should be conformally invariant. This leads to the relation between $\Delta_1$
and the scaling dimension of the operator ${\it O}$, dual to ${\tilde{\Delta}}$,
which reads;
\begin{equation}
d-\tilde{\Delta}= \Delta_2.
\end{equation}
One can check that $\tilde{\Delta}={\Delta_1}$.
As one can see this is an irrelevant operator i.e.,
it corresponds to a deformations which is irrelevant in the IR but
important in the UV. This is consistent with fact that the bulk deformation is a higher derivative term suppressed by the string length scale.

It may be noted that these
CFT operators scale as  $\Delta_1\sim N^{2/3}$ in five dimensions, or
$\Delta\sim N^{1/4}$ in ten dimensions. It has been suggested that the  high energy  excitation in the bulk will correspond to the
CFT operators scaling as  $\Delta_1\sim N^{2/3}$ in five dimensions, or
$\Delta\sim N^{1/4}$ in ten dimensions \cite{1c}. Thus, the   corrections generated from the generalized uncertainty principle
actually correspond to
high energy excitations in the bulk.
This is also what is expected from a effective field theory perspective,
as we are studding low energy effective phenomena. The low energy effective  field theory equations are obtained
by integrating the high energy excitations away. Thus,   next to the leading order corrections to
classical action for the low energy effective field theory are obtained by integrating the minimum measurable length scale out.
 Thus, we have identified the deformations of the
 CFT operators scaling as $\Delta_1\sim N^{2/3}$ in five dimensions, or
$\Delta\sim N^{1/4}$ in ten dimensions with   corrections in the bulk theory which were generated from the generalized uncertainty principle.

\section{Holographic Renormalization}

In this section, we will calculate  correlation  functions in this field theory deformed by generalized uncertainty principle.
In fact, just from the  conformal invariance, we can predict the form of these functions,
 \begin{equation}\label{twopointfunction}
  \langle O(x)O(0)\rangle= {C \over |x|^{2\tilde{\Delta}}}.
 \end{equation}
However, to derive an explicit form for these functions, we need to apply the techniques of holographic renormalization \cite{4}-\cite{4a}
  to the massive scalar in $AdS$ with Planck scale deformation.
We write the boundary value of any bulk field $\Phi(z,x)$ as,
\begin{equation}
\Phi_0(x)=\Phi(z=0,x)=\Phi |_{\partial\rm{AdS}}(x),
\end{equation}
where $\Phi_0(x)$ is a source of a dual operator $O$ in the CFT side. The generating functional of this CFT can be written as
\begin{equation}
Z_{\rm{CFT}}[\Phi_0]=\Big\langle \exp\Big[\int \Phi_0 \mathcal{O} \Big]  \Big\rangle=Z_{\rm{gravity}}[\Phi\rightarrow \Phi_0],
\end{equation}
where the path integral in $Z_{\rm{gravity}}[\Phi\rightarrow \Phi_0]$
is over all fields  whose boundary value is    $\Phi_0$.
In the limit,  where classical gravity dominates  the partition function $Z_{\rm{gravity}}[\Phi\rightarrow \Phi_0]$, can be approximated by
\begin{equation}
Z_{\rm{gravity}}[\Phi\rightarrow \Phi_0]\approx e^{S_{\rm{on-shell}} }.
\end{equation}
 The on-shell gravity action in $AdS$ suffers from divergences due
 to the infinite volume of $AdS$
 and one needs to replace the action by a renormalized version $S_{\rm{on-shell}}^{\rm{ren}}$.
 The two-point function (\ref{twopointfunction}) will be given by
 \begin{equation}
 \langle O(x) O(0)\rangle=\frac{\delta^2 S_{\rm{on-shell}}^{\rm{ren}}[\Phi\rightarrow \Phi_0] }{\delta \varphi(x) \delta \varphi(0) },
 \end{equation}
 where
 \begin{equation}
 \varphi(x)=\lim_{z\mapsto 0 }z^{\Delta-d} \Phi(z,x).
 \end{equation}

We follow the methods used in \cite{5},  and use the following form for the metric of $AdS_{d+1}$
\begin{equation}\label{eq:metric}
ds^2=G_{MN}dx^M dx^N=\frac{d\rho^2}{4\rho^2}+\frac{1}{\rho} dx^i dx^i,
\end{equation}
where the $AdS$ radius has been set equal to one.
The Laplacian in this metric is given by
\begin{equation}
\Box=(-2d+4)\rho \partial_\rho+4\rho^2 \partial^2_\rho+\rho\Box_0,
\end{equation}
where $\Box_0=\delta^{ij}\partial_i \partial_j$.
In the metric given by Eq. (\ref{eq:metric}), we have
\begin{equation}
n^{\mu}=(n^\rho,0,0,0),\qquad \gamma_{ij}=\frac{\delta_{ij}}{\rho},\quad \sqrt{\gamma}=\rho^{-\frac{d}{2}},
\quad n^\rho=\frac{1}{\sqrt{G_{\rho\rho}}}=2\rho.
\end{equation}
The equation of motion of the deformed scalar in $AdS$ is
\begin{equation}\label{deformedscalar}
\Big(\Box-m^2-\frac{1}{M_P^2}\Box^2\Big)\Phi=0.
\end{equation}
We look for solutions of this equation of the form
\begin{equation}
\Phi(\rho,x)=\rho^{(\Delta-d)/2}\phi(\rho,x),\qquad \phi(\rho,x)=\phi_{(0)}(x)+\rho\phi_{(2)}(x)+\rho^2\phi_{(4)}(x)+\cdots.
\end{equation}
Now we use  this form of  $\Phi(\rho,x)$, and recursively solve the equation at each order of $\rho$.
So, at order $\rho^0$,  one gets
\begin{equation}
\Delta  (d-\Delta ) (\Delta  (d-\Delta )+M)+m^2 M=0
\end{equation}
which is the relation between the mass and conformal dimension given in Eq. (\ref{massVscd}).
At higher orders  one gets
\begin{eqnarray}
\phi_{(2)}&=&\frac{\Box_0 \phi_{(0)}}{2(2\Delta-d-2)}, \nonumber \\[6pt]
\phi_{(4)}&=&\frac{\Box_0 \left(\Box_0 \phi (0)-(2 d (\Delta -3)-2 ((\Delta -6) \Delta +10)+M)\phi_{(2)} \right)}{2 (d-2 \Delta +4)
(2 d (\Delta -2)-2 ((\Delta -4) \Delta +8)+M)},\nonumber \\[6pt]
\phi_{(6)}&=&\frac{\Box_0 \left(2 \Box_0 \phi_{(2)}- \left(2 d (\Delta -5)-2 \Delta ^2+20 \Delta
+M-52\right)\phi_{(4)}\right)}{2 (d-2 \Delta +6) (2 d (\Delta -3)-2 ((\Delta -6) \Delta +18)+M)}\nonumber \\
&&\ \,  \vdots.
\end{eqnarray}
If we write everything in terms of $\phi_{(0)}$, we get
\begin{equation}
\phi_{(2N)}=
\frac{\Box_0^N \phi_{(0)}}{\prod_{n=1}^N 2n(2\Delta-d-2n)}.
\end{equation}
This is the same form one gets in the undeformed case.

Now we evaluate the on-shell action on the classical solution to read off the counter terms.
First, we consider operators for which $\Delta\neq d/2+k$
\begin{eqnarray}
S_b&=&\frac{1}{2} \int d^{d}x \sqrt{\gamma}\,  n^\mu \Big[
\Phi\partial_\mu\Phi+\beta   \Big(\partial_\mu\Phi \Box\Phi-\Phi\partial_\mu\Box\Phi \Big)
 \Big]\nonumber
\\[6pt]
&=&\frac{1}{2} \int d^{d}x \sqrt{\gamma}\,  n^\rho \Big[
\Phi\partial_\rho\Phi+\beta   \Big(\partial_\rho\Phi \Box\Phi-\Phi\partial_\rho\Box\Phi \Big)
 \Big].\nonumber
\\[6pt]
&=& \int_{\rho=\epsilon} d^{d}x\, \Big(\epsilon^{\frac{d}{2}-\Delta}a_{(0)}+\epsilon^{\frac{d}{2}-\Delta+1}a_{(2)}+
\epsilon^{\frac{d}{2}-\Delta+2}a_{(3)} \cdots\Big).
\end{eqnarray}
where
\begin{eqnarray}
a_{(0)}&=&\frac{1}{2} (d-\Delta )\phi_0^2,\nonumber \\[5pt]
a_{(2)}&=& -\beta  \phi_0\Box_0 \phi_0
+\frac{\big(d-\Delta +1+\beta (4-4\Delta+2d)\big) \phi_0 \Box_0 \phi_0}{2 (2 \Delta-d -2)},\nonumber \\[5pt]
a_{(2)}&=&\frac{(d-\Delta +1)}{2 (2 \Delta-d -2)} \phi_0\Box_0 \phi_0,\nonumber \\
a_{(3)}&=&\frac{(d-\Delta +3) (d-2\Delta +5)}{12 (d-2 \Delta +2)^2 (d-2 \Delta +4) (d-2 \Delta +6)} \phi_0\Box_0^3 \phi_0.
\end{eqnarray}
These coefficients are of the same form like in the undeformed case, i.e.,  the higher derivative terms in the
boundary action do not introduce new divergences,  and so,  we do not need any new counter terms.
Now up to second order,  we can write \cite{5}
\begin{eqnarray}
\phi_{(0)}&=&\epsilon^{-(d-\Delta)/2}\bigg(\Phi(x,\epsilon)-\frac{1}{2(2\Delta-d-2)}\Box_{\gamma} \Phi(x,\epsilon) \bigg)\nonumber \\[4pt]
\phi_{(2)}&=&\epsilon^{-(d-\Delta)/2-1}\frac{1}{2(2\Delta-d-2)}\Box_{\gamma} \Phi(x,\epsilon).
\end{eqnarray}
From the above expansions one can write down the counter term action for operators with $\Delta\neq d/2+1$
\begin{equation}
S_{\rm{ct}}=-\int d^dx \sqrt{\gamma}\bigg(\frac{d-\Delta}{2}\Phi^2 +\frac{1}{2(2\Delta-d-2)}\Phi\Box_\gamma\Phi\bigg)
\end{equation}
The complete regularized action is now given by
\begin{eqnarray}
S&=&\frac{1}{2} \int d^{d}x \sqrt{\gamma}\,  n^\mu \Big[
\Phi\partial_\mu\Phi+\beta \Big(\partial_\mu\Phi \Box\Phi-\Phi\partial_\mu\Box\Phi \Big)\nonumber \\[3pt]
&&\qquad- \frac{d-\Delta}{2}\Phi^2 -\frac{1}{2(2\Delta-d-2)}\Phi\Box_\gamma\Phi
 \Big].
\end{eqnarray}

For $\Delta=d/2+k$, we will need to introduce logarithmic term in the expansion of $\Phi(\rho,x)$.
We consider $\Delta=d/2+1$, where the expansion takes the form
\begin{equation}
\Phi(\rho,x)=\phi_{(0)}+\rho\big(\phi_{(2)}+\log \rho \psi_{(2)})+\cdots.
\end{equation}
By plugging this into Eq. (\ref{deformedscalar}),  we find
\begin{equation}
\psi_{(2)}=-\frac{2+\beta (d^2-4)}{8+\beta(d^2-4)} \Box_0 \phi_{(0)},
\end{equation}
Since the coefficient of the logarithmic term gives the matter conformal anomaly for the usual case,  the
higher derivative deformation of the scalar field  theory is expected to contributes to the matter conformal anomaly.
Thus, the matter conformal anomaly is generated by deforming the scalar field theory in the bulk by generalized uncertainty
principle.
The new contribution is of order $1/N^{1/2}$.

For an operator with $\Delta= d/2+1$, the counter term action takes the form
\begin{equation}
S_{\rm{ct}}=-\int d^dx \sqrt{\gamma}\bigg(\frac{d-\Delta}{2}\Phi^2 -\frac{2+\beta(d^2-4)}{8+\beta (d^2-4)}
\log \epsilon \Phi \Box_\gamma \Phi \bigg).
\end{equation}
The renormalized  on-shell regularized action with counter terms is now given by
\begin{eqnarray}
S&=& \int  d^{d}x\, \sqrt{\gamma} n^\mu \Big[
\Phi\partial_\mu\Phi+\beta   \Big(\partial_\mu\Phi \Box\Phi-\Phi\partial_\mu\Box\Phi \Big)\nonumber \\[4pt]
 &&\qquad- \frac{d-\Delta}{2}\Phi^2 -\frac{2+\beta (d^2-4)}{8+\beta (d^2-4)}  \log \epsilon \Phi \Box_\gamma \Phi\bigg).
\end{eqnarray}
The above actions can be easily used to compute  correlation functions \cite{5}. For example, the
  the one-point function can be written as
\begin{equation}
\langle O_\Phi \rangle =
\lim_{\epsilon\mapsto 0}\bigg(\frac{1}{\epsilon^{\Delta/2}}\frac{1}{\sqrt{\gamma}} \frac{\delta S}{\delta \Phi(x,\epsilon)}\bigg).
\end{equation}
Thus, we are able to find any correlation functions using this on-shell regularized action. It may be noted that by using holographic  renormalization, it was   demonstrated that the
deformation of the scalar field  theory based on the generalized uncertainty principle
contributes to the matter conformal anomaly.

\section{Conclusion}

In this paper, we deformed a  free massive scalar field theory on $AdS$ by generalized uncertainty principle. It was demonstrated that the higher derivative terms produced from derivative expansion in effective field theory exactly matches the deformation produced by the generalized uncertainty principle. This was because the
derivative expansion of the effective field theory was obtained by integrating out the scale corresponding to the minimum measurable length.
We also explicitly calculated the boundary field theory dual to this scalar field theory with higher derivative corrections.
Furthermore, it was demonstrated that
higher derivatives  correspond to the existence of two massive scaling dimensions.
Finally, we calculated the correlation functions using holographic renormalization.
Thus, the
 UV divergences of the correlation functions on the boundary were
renormalized. These  UV divergences  were related  to the IR divergences on the bulk. In fact,
the IR divergences on the bulk are  the same as near-boundary effects,
and so, they were dealt by  using holographic renormalization.
Using holographic  renormalization, it was shown that the
 deformation of the scalar field  theory by the generalized uncertainty principle
 contributes to the matter conformal anomaly.
It was also found that the new
CFT operators scaled as  $\Delta_1\sim N^{2/3}$ in five dimensions, or
$\Delta\sim N^{1/4}$ in ten dimensions. As it had been suggested that the purely stringy excitation in the bulk will correspond to the
CFT operators scaling as  $\Delta_1\sim N^{2/3}$ in five dimensions, or
$\Delta\sim N^{1/4}$ in ten dimensions \cite{1c}, we concluded that these  higher derivative corrections may actually correspond to
high energy excitation in the bulk. In fact, we also argued that this is what is expected to occur from
an effective field theory perspective, because we were
  studding low energy effective phenomena. As the low energy effective  field theory equations were obtained
by integrating the high energy  excitation away, the next to the lead order corrections to
classical action for the low energy effective field theory would appear as a higher derivative correction.
Thus, we   identified the deformations of the
 CFT operators scaling as $\Delta_1\sim N^{2/3}$ in five dimensions, or
$\Delta\sim N^{1/4}$ in ten dimensions, with higher derivative corrections in the bulk theory.

It may be noted that we only analyzed the higher derivative corrections for an ordinary scalar field theory
on $AdS$ and related it to the conformal field theory on its boundary. The precise correspondence between a
ordinary scalar field theory on $AdS$ and
a suitable conformal field theory on its boundary is given by the Rehren duality   \cite{a1}-\cite{b1}.
It would thus be interesting to analyze the the boundary dual to the scalar field theory on
the bulk in the framework of algebraic holography. It may also be interesting to
analyze the bulk action of various supergravity theories in the framework of effective field theories.
We will expect that the bulk action will receive higher derivative corrections from purely stringy excitations.
Then it will be possible to relate these higher derivative corrections for the bulk supergravity action
  to the superconformal field theories on the boundary.
As the full string theory is dual to boundary superconformal field theory, we expect that
the  conformal dimension of marginal operators will not  receive any correction from these
 purely stringy excitations.
However,
conformal dimensions of both the relevant and the irrelevant operators are expected to
receive corrections.
It would be interesting to perform this analysis explicitly, and demonstrate this to be the case.

\section*{Acknowledgments}
We would like to thank Adel Awad for useful discussions, and pointing out to us the appearance of the second massive
scale in the theory. The research of Ali Nassar is supported by the STDF project 13858.

\end{document}